\def\breakon{\end{multicols}\widetext\vspace{-.6cm}
\noindent\rule{.49\linewidth}{.3mm}\rule{.3mm}{.5cm}\vspace{0.0cm}}
\def\breakoff{\vspace{-0.45cm}
\noindent
\rule{.50\linewidth}{.0mm}\rule[-.47cm]{.3mm}{.5cm}\rule{.49\linewidth}{.3mm}
\vspace{-0.25cm}
\begin{multicols}{2}   }
\documentstyle[aps,epsf,multicol]{revtex}
\begin{document}

%%%%%%%%%%%%%%%%%%%%%%%%%%%%%%%%%%%%%%%%%%%%%%%%%%%%%%%%%%
%% multicol commands
%%%%%%%%%%%%%%%%%%%%%%%%%%%%%%%%%%%%%%%%%%%%%%%%%%%%%%%%%%
% This is needed so that figure captions come out right 
% when using the "multicols" setting.
\makeatletter
\renewenvironment{table}
  {\let\@capwidth\linewidth\def\@captype{table}}
  {}

\renewenvironment{figure}
  {\let\@capwidth\linewidth\def\@captype{figure}}
  {}
\makeatother
%%
%%%%%%%%%%%%%%%%%%%%%%%%%%%%%%%%%%%%%%%%%%%%%%%%%%%%%%%%%%

\title{Kondo Effect in Quantum Dots Coupled to 
 Luttinger Liquid Leads}
\author{Eugene H. Kim}
\address{Department of Physics and Astronomy, 
         McMaster University, Hamilton, Ontario, Canada L8S-4M1 }
\maketitle

\begin{abstract}
We consider the Kondo effect in quantum dots coupled to Luttinger 
liquid leads, focussing on the case of repulsive interactions and 
spin SU(2) symmetry in the leads.  We find that the system can
flow to the 1-channel or 2-channel Kondo fixed points, depending
on the interactions in the system.  We compute the conductance and 
find that the qualitative behavior is strongly dependent on the 
interactions.  Finally, we point out a consequence of 2-channel 
Kondo physics, which should be observable in thermal conductance 
measurements.
\end{abstract}

%%%%%%%%%%%%%%%%%%%%%%%%%%%%%%%%%%%%%%%%%%%%%%%%%%%%%%%%%%%%%%%%%%%%%%%%%%%
%%%%%%%%%%%%%%%%%%%%%%%%%%%%%%%%%%%%%%%%%%%%%%%%%%%%%%%%%%%%%%%%%%%%%%%%%%%

\vspace{.15in}
\begin{multicols}{2}

The Kondo effect, which deals with a single magnetic 
impurity in a sea of conduction electrons, has received
an enormous amount of attention over the years.  Kondo's
original work over thirty-five years ago was intended 
to explain the anomalous resistivity observed in magnetic 
alloys.\cite{kondo}  Since then, a variety of variations
of the Kondo Hamiltonian have been introduced, and some
interesting manifestations of Kondo physics have been 
suggested.\cite{cox}  
One of the most recent manifestation of the Kondo effect 
has been provided by quantum dots.\cite{dots}  In quantum 
dots with an odd number of electrons, resonant transmission 
through the dot is observed due to the Kondo effect.\cite{ng}  
The quantum dot provides a fascinating place to study Kondo 
physics because there are many parameters which can be controlled
in an experiment.  Therefore, many aspects of the Kondo effect
can be probed.

\vspace{.2in}
\begin{figure}
\epsfxsize=3.0in
\centerline{\epsfbox{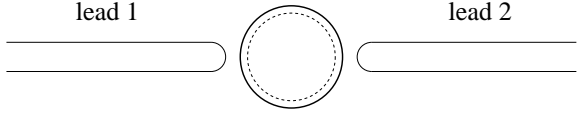} }
\vspace{.2in}
\caption{ The model: A quantum dot coupled to Luttinger liquid
 leads. }
\label{fig:setup}
\end{figure}
\vspace{.2in}

In most treatments of the Kondo effect in quantum dots, the 
leads were taken to be Fermi liquids and interactions in the 
leads were ignored.  This is sufficient if the leads are two 
or three dimensional electron gases, where interactions affect 
the low energy properties only perturbatively.  However, 
in one dimension, arbitrarily weak interactions completely
modify the ground state; the low energy excitations are
described by a Luttinger liquid rather than a Fermi 
liquid.\cite{haldane}  In contrast to a Fermi liquid,
the low energy excitations in a Luttinger liquid are
spin and charge density fluctuations; there are no well
defined single particle excitations.  Moreover, the speeds
of the spin and charge density fluctuations are different
i.e. spin and charge separate.
In this work, we consider the Kondo effect in quantum dots 
coupled to Luttinger liquid leads.  The setup we consider is 
shown in Fig.~\ref{fig:setup}.  In what follows, we focus on 
the case of repulsive interactions in the leads and assume 
the system to have spin SU(2) symmetry.

%%%%%%%%%%%%%%%%%%%%%%%%%%%%%%%%%%%%%%%%%%%%%%%%%%%%%%%%%%%%%%%%%%%%%%%%%%%
%%%%%%%%%%%%%%%%%%%%%%%%%%%%%%%%%%%%%%%%%%%%%%%%%%%%%%%%%%%%%%%%%%%%%%%%%%%
  
To model the dot, we consider what has come to be the canonical
model for small quantum dots --- the Anderson impurity model.  
The Hamiltonian, including the coupling to the leads, is
\begin{eqnarray}
 H_{\rm dot} & = & \varepsilon_0 \sum_s 
    d^{\dagger}_s d^{\phantom \dagger}_s 
  + U_0~ d^{\dagger}_{\uparrow}d^{\phantom \dagger}_{\uparrow}
         d^{\dagger}_{\downarrow} d^{\phantom \dagger}_{\downarrow}  
 \nonumber \\
  & + & t' \left[\left(\psi^{\dagger}_{1,s}(0) 
    + \psi^{\dagger}_{2,s}(0) \right)d_s + h.c. \right]  \, ,
\end{eqnarray}
where $\psi^{\dagger}_{i,s}$ creates an electron with spin-$s$
in lead-$i$; $d^{\dagger}_s$ creates an electron with spin-$s$ 
on the dot; $\varepsilon_0$ is the energy level of the dot, 
which can be controlled by a gate voltage; $U_0$ is the 
charging energy; $t'$ is the tunneling matrix element between 
the leads and the dot.  In this work, we will focus on the case
where the tunneling matrix elements of both leads are equal.
Being interested in the Kondo regime of the dot, we integrate 
out the charge fluctuations on the dot.  Working to second order 
in perturbation theory, the following interactions are generated
\begin{eqnarray}
 H_{\rm int} & = & J~ \tau \cdot \frac{{\bf \sigma}_{s,s'}}{2}
     \left(\psi^{\dagger}_{1,s}(0) \psi^{\phantom \dagger}_{1,s'}(0)
   + \psi^{\dagger}_{2,s}(0) \psi^{\phantom \dagger}_{2,s'}(0) 
   \right)  \nonumber \\
 & + & J'~ \tau \cdot \frac{{\bf \sigma}_{s,s'}}{2}
    \left(\psi^{\dagger}_{1,s}(0) \psi^{\phantom \dagger}_{2,s'}(0)
   + h.c. \right)  \nonumber \\  
 & + & V~ \left(\psi^{\dagger}_{1,s}(0) \psi^{\phantom \dagger}_{2,s}(0)
   + h.c. \right)  \nonumber \\ 
 & + & V'~ \left(\psi^{\dagger}_{1,s}(0) \psi^{\phantom \dagger}_{1,s}(0)
   + \psi^{\dagger}_{2,s}(0) \psi^{\phantom \dagger}_{2,s}(0) \right)
   \, .
\label{schreifferwolff}
\end{eqnarray}
The coupings can be computed for a given microscopic model of the 
leads (e.g. Hubbard model); we will take them as phenomenological 
parameters.  However, it is important to note that $J > 0$ 
and $J' > 0$.  

Being interested in the low energy properties of the system, 
we expand the electron operators in the leads in terms of right 
and left movers
\begin{equation}
 \psi_{i,s}(x) = e^{ik_F x} \psi_{R,i,s}(x) 
    + e^{-ik_F x} \psi_{L,i,s}(x)  \, ,
\label{linearizedelectron}
\end{equation}
where $k_F$ is the Fermi wavevector, and $\psi_{R,i,s}$ and 
$\psi_{L,i,s}$ are the (slowly varying) right and left moving 
fermion operators.
In bosonized form, the Hamiltonian describing low energy 
fluctuations in the leads is\cite{backscattering}
\begin{eqnarray}
 H & = & \frac{v_{\rho}}{2} \sum_{i=1}^2 \int_{-\infty}^0 dx~ 
    K~ \pi_{i,\rho}^2 + \frac{1}{K}~(\partial_x \phi_{i,\rho} )^2 
 \nonumber \\ 
   & + & \frac{v_{\sigma}}{6\pi} \sum_{i=1}^2
   \int_{-\infty}^0 dx~ {\bf J}_{R,i}^2 + {\bf J}_{L,i}^2  \, ,
\label{leadhamiltonian}
\end{eqnarray}
where $i=1,2$ labels the leads; $\phi_{i,\rho}$ and $\pi_{j,\rho}$
are canonically conjugate fields describing charge fluctuations
and satisfy
$[\phi_{i,\rho},\pi_{j,\rho}] = i\delta_{i,j}\delta(x_1 - x_2)$;
${\bf J}_{R,i}$ and ${\bf J}_{L,i}$ are currents satisfying
the SU(2) Kac-Moody algebra at level-1 and describe spin fluctuations.
The Luttinger parameter, $K$, is determined by the 
interactions in the system --- $K < 1$ for repulsive 
interactions, $K > 1$ for attractive interactions, and
$K=1$ for a non-interacting system.  
At the ultraviolet fixed point, the leads are decoupled 
from the dot.  Therefore the right and left moving fermion 
fields satisfy
\begin{equation}
 \psi_{R,i,s}(x=0) = \psi_{L,i,s}(x=0) \, .
\label{dirichlet}
\end{equation}
Eq.~\ref{dirichlet} implies that $\psi_{L,i,s}(x) = \psi_{R,i,s}(-x)$.
Hence, $\psi_{L,i,s}$ may be regarded as the analytic continuation
of $\psi_{R,i,s}$ to positive x, and we can write our Hamiltonian 
solely in terms of right movers.\cite{eggert,fabrizio} 

The Hamiltonian we consider (Eq.~\ref{schreifferwolff} and
Eq.~\ref{leadhamiltonian}) is related to the problem of
a magnetic impurity in a Luttinger liquid, which has been
considered in several works.\cite{fabrizio,luttingerkondo}  
In that context, this Hamiltonian was first considered in 
Ref.~\ref{fabrizio} to describe a magnetic impurity in a 
Luttinger liquid with a strong scattering potential at the 
impurity site.  There, they identified that the system could 
flow to the 1-channel or 2-channel Kondo fixed points, depending 
on the interactions in the system.  In this work, our main goal 
is understanding transport through a quantum dot.  More 
specifically, how do the various fixed points which arise 
influence the transport through the dot?  

%%%%%%%%%%%%%%%%%%%%%%%%%%%%%%%%%%%%%%%%%%%%%%%%%%%%%%%%%%%%%%%%%%%%%%%%%%%
%%%%%%%%%%%%%%%%%%%%%%%%%%%%%%%%%%%%%%%%%%%%%%%%%%%%%%%%%%%%%%%%%%%%%%%%%%%

Before we proceed with the case of Luttinger liquid leads, it 
is useful to review the case of non-interacting leads.  For 
non-interacting leads, one finds that $J = J'$ in 
Eq.~\ref{schreifferwolff}.  Also, the $V$ and $V'$ terms 
are usually ignored, since their effects are very small.
Then, by working in the ``even-odd'' basis,
$\psi_{e/o,s} = (1/\sqrt{2})(\psi_{1,s} \pm \psi_{2,s})$,
the Hamiltonian becomes $H = H_0 + H_{\rm int}$, 
where $H_0$ is the Hamiltonian for free electrons and
\[
 H_{\rm int} = J~ {\bf \tau} \cdot \frac{{\bf \sigma}_{s,s'}}{2}
 \psi^{\dagger}_{e,s} \psi^{\phantom \dagger}_{e,s'}  \, .
\]
We see that only the even channel couples to the impurity,
giving us a 1-channel Kondo model; the odd channel remains
free.  The low energy (i.e. 1-channel Kondo) fixed point 
is given by an electron in the even channel bound in a singlet 
to the electron on the dot.  Resonant transmission occurs 
because electrons can tunnel freely through the odd channel.
Essentially, the two semi-infinite leads have been joined to 
form one infinite lead, due to the Kondo effect.

%%%%%%%%%%%%%%%%%%%%%%%%%%%%%%%%%%%%%%%%%%%%%%%%%%%%%%%%%%%%%%%%%%%%%%%%%%%
%%%%%%%%%%%%%%%%%%%%%%%%%%%%%%%%%%%%%%%%%%%%%%%%%%%%%%%%%%%%%%%%%%%%%%%%%%%

For the case of Luttinger liquid leads, the even-odd basis
is not useful, because the Hamiltonian becomes non-local.
Therefore, we will proceed along different lines and 
change variables in a different way.  For the spin degrees
of freedom, we work in terms of a total spin variable, 
${\bf J}_R = {\bf J}_{R,1} + {\bf J}_{R,2}$, and fields 
describing the relative spin fluctuations between the two 
leads.  To deduce the fields describing the relative spin
fluctuations, we note that ${\bf J}_R$ are currents of the 
SU(2) WZW model at level-2; the WZW model at level-2 has
a central charge of 3/2.  Our original description was 
in terms of the ${\rm SU(2)}_1 \times {\rm SU(2)}_1$ WZW 
model with central charge 2.  Therefore, the relative spin 
fluctuations must account for a missing central charge of 
1/2.  This is precisely the central charge of the Ising
model.  Therefore, we can describe the spin degrees of
freedom in terms of an SU(2) WZW model at level-2 for the 
total spin and an Ising model for the relative spin fluctuations 
between the two leads.\cite{goddard,ludwig2,frojdh}  
The spectrum of the WZW model at level-2 is classified by 
its 3 primary fields --- the spin-0 field $I$ of dimension 
0, (the identity operator); the spin-1/2 field $g_s$ of 
dimension 3/16; and the spin-1 field ${\bf \phi}$ of 
dimension 1/2.  The spectrum of the Ising model is classified 
by its 3 primary fields --- $I$ of dimension 0 (the identity 
operator); the order parameter field $\sigma$ of dimension 
1/16; and the energy field $\epsilon$ of dimension 1/2.  The 
WZW fields satisfy the operator product expansions (OPE's)
\begin{equation} 
 [g] \times [g] \rightarrow [I] + [{\bf \phi}]  \ \ , \ \ 
 [{\bf \phi}] \times [{\bf \phi}] \rightarrow [I]  \ \ , \ \
\label{WZWOPE}
\end{equation}
and the Ising fields satisfy 
\begin{equation}
 [\sigma] \times [\sigma] \rightarrow [I] + [\epsilon]  \ \ , \ \ 
 [\epsilon] \times [\epsilon] \rightarrow [I]   \, .
\label{IsingOPE}
\end{equation} 
In terms of the charge, WZW, and Ising fields, the electron 
operator is
\begin{equation}
 \psi_{R,i,s} \sim e^{i\sqrt{2\pi}\phi_{R,i}} g_s~ \sigma  \, .
\label{cosetelectron}
\end{equation}
It will also prove useful to form the combinations
\[
 \phi_{R,c/f} = \frac{1}{\sqrt{2}}\left(\phi_{R,1,\rho} 
            \pm \phi_{R,2,\rho}\right)  \, 
\]
for the charge degrees of freedom.  $\phi_{R,c}$ is a {\sl charge} 
field describing total charge fluctuations of the system; 
$\phi_{R,f}$ is a {\sl flavor} field describing the relative 
charge fluctuations between the two leads.  

%%%%%%%%%%%%%%%%%%%%%%%%%%%%%%%%%%%%%%%%%%%%%%%%%%%%%%%%%%%%%%%%%%%%%%%%%%%
%%%%%%%%%%%%%%%%%%%%%%%%%%%%%%%%%%%%%%%%%%%%%%%%%%%%%%%%%%%%%%%%%%%%%%%%%%%

Let us first (formally) consider the case where the system
has particle hole symmetry.  In this case the $V$ and $V'$ 
terms are absent from Eq.~\ref{schreifferwolff}.  Using 
Eq.~\ref{cosetelectron} and the OPE's of Eq.~\ref{WZWOPE} and 
\ref{IsingOPE}, we obtain 
\begin{equation}
 H_{\rm int} = v_{\sigma}~ \lambda_1~ {\bf \tau} \cdot {\bf J}_R(0)
   + \sqrt{v_{\sigma} v_{\rho}}~ \lambda_2~ {\bf \tau} \cdot {\bf \phi}~ 
     \cos \sqrt{4\pi} \phi_{R,f}(0)  \, ,
\label{bosonschreifferwolff1}
\end{equation}
where $\lambda_1$ and $\lambda_2$ are dimensionless couplings, with
$v_{\sigma}~ \lambda_1 \sim J$ and 
$\sqrt{v_{\sigma}v_{\rho}}~\lambda_2 \sim J'$.
We begin by considering the effects of 
Eq.~\ref{bosonschreifferwolff1} on the ultraviolet fixed 
point via a renormalization group (RG) analysis.  To second 
order in the couplings,\cite{cardy} the RG equations for the 
parameters are
\begin{equation}
 \frac{d\lambda_1}{dl} = \lambda_1^2 + \lambda_2^2 \ \ , \ \ 
 \frac{d\lambda_2}{dl} = 
    \frac{1}{2}\left(1 - \frac{1}{K} \right) \lambda_2 
  + 2~ \lambda_1 \lambda_2  \, .  
\label{ultravioletRG}
\end{equation}
A few words are in order about the RG equations.  At the 
ultraviolet fixed point, ${\bf \tau}\cdot {\bf J}_{R}$ is 
marginal.  Since $\lambda_1 > 0$, quantum corrections push 
${\bf \tau} \cdot {\bf J}_{R}$ to be marginally relevant.  
On the other hand, the interlead tunneling term is irrelevant 
for repulsive interactions ($K < 1$).  Therefore, $\lambda_2$ 
initially decreases under the RG.  However, from 
Eq.~\ref{ultravioletRG} we see that the growth of $\lambda_1$ 
will also cause $\lambda_2$ to eventually grow.  For $K \approx 1$, 
$\lambda_2$ will grow almost immediately.  
Similar to the case of non-interacting electrons, the system 
flows to the 1-channel Kondo fixed point.  However, for $K$ 
sufficiently smaller than unity, we can have $\lambda_2 \ll 1$ 
while $\lambda_1$ has already grown to ${\cal O}(1)$.  If 
$\lambda_2 = 0$, we would have a 2-channel Kondo model, which 
is known to have a nontrivial ${\cal O}(1)$ fixed point.  For 
the case $\lambda_2 \ll 1$ with $\lambda_1 = {\cal O}(1)$, the 
system flows close to the 2-channel Kondo fixed point.  In that 
case, it is appropriate to consider the behavior near the 2-channel 
Kondo fixed point with $\lambda_2$ as a perturbation.

%%%%%%%%%%%%%%%%%%%%%%%%%%%%%%%%%%%%%%%%%%%%%%%%%%%%%%%%%%%%%%%%%%%%%%%%%%%

The 2-channel Kondo fixed point occurs for $\lambda_1^* = 1/2$.  
For this value of the coupling, we can define new spin currents, 
${\bf J}'_R(x) = {\bf J}_R(x) + 2\pi {\bf \tau}\delta(x)$,
which satisfy the same Kac-Moody algebra as the original 
spin currents.  The impurity spin has now disappeared from 
the problem, leaving behind a new boundary condition.\cite{ludwig1}  
The spectrum at the 2-channel Kondo fixed point is obtained 
by fusion with spin-1/2; the operator content is obtained by 
double fusion.\cite{ludwig1}  Upon double fusion, 
\begin{equation}
 [I] \rightarrow [I] + [{\bf \phi}] \ \ , \ \ 
 [g] \rightarrow [g]  \ \ , \ \
 [{\bf \phi}] \rightarrow [I] + [{\bf \phi}]  \, .
\label{doublefusionrules}
\end{equation}
To understand the properties near the 2-channel Kondo fixed 
point, we must consider the various operators which are allowed.  
Since the system has spin SU(2) symmetry, any operator from
the spin sector must transform as a singlet.  This forces
${\bf J}_{-1} \cdot {\bf \phi}(0)$, the first descendant of
${\bf \phi}$ with dimension 3/2, as the leading irrelevant 
operator from the spin sector.  In the ordinary 2-channel 
Kondo problem, only the spin sector enters; in this system, 
the other sectors contribute as well.  Operators from the 
charge sector are not allowed, since the total charge of 
the system is conserved.  Also, operators from the Ising 
sector are not allowed, since the Hamiltonian is invariant 
with respect to lead interchange --- lead 1 $\leftrightarrow$ 
lead 2.\cite{frojdh}  However, operators from the flavor sector 
are allowed, due to the $\lambda_2$ term in 
Eq.~\ref{bosonschreifferwolff1}.  The allowed operators can 
be deduced by double fusion.  Using Eq.~\ref{doublefusionrules}, 
we find that 
\begin{equation}
\cos \sqrt{4\pi} \phi_{R,f}(0)  
\label{2channeltunneling}
\end{equation}
is allowed.  This operator has dimension $1/2K$; it is relevant 
for $K > 1/2$ and irrelevant for $K < 1/2$.   

%%%%%%%%%%%%%%%%%%%%%%%%%%%%%%%%%%%%%%%%%%%%%%%%%%%%%%%%%%%%%%%%%%%%%%%%%%%

For $K < 1/2$, the operator in Eq.~\ref{2channeltunneling} is 
irrelevant, and the 2-channel Kondo fixed point is stable. 
However, for $K > 1/2$, this operator is relevant and induces 
a flow away from the 2-channel Kondo fixed point.  The boundary 
conditions at the ultraviolet fixed point (Eq.~\ref{dirichlet}) 
imply that the flavor sector satisfies Dirichlet boundary 
conditions; near the 2-channel Kondo fixed point, the flavor 
sector continues to satisfy Dirichlet boundary conditions.  The 
operator in Eq.~\ref{2channeltunneling} induces a flow from 
Dirichlet to Neumann boundary conditions.  This operator is 
similar to what appears in the case of the anisotropic 2-channel 
Kondo model with a spin-1/2 impurity.\cite{ludwig3}  There, it 
is known that channel anisotropy is a relevant perturbation to 
the 2-channel Kondo fixed point and causes the system to ultimately
flow to the 1-channel Kondo fixed point.  Physically, this occurs 
because the channel with the larger exchange coupling screens the 
impurity.\cite{blandin}  Here, the physics is similar.  The operator 
in Eq.~\ref{2channeltunneling} induces a flow from the 2-channel 
to the 1-channel Kondo fixed point (for $K > 1/2$), and the 
stronger channel screens the impurity.  As in the case of 
non-interacting leads, the stronger channel is the ``even'' 
channel, $\psi_{e,s} =(1/\sqrt{2})(\psi_{1,s} + \psi_{2,s})$.  
 
We see that for $K > 1/2$, the system ultimately flows to the
1-channel Kondo fixed point.  We must now consider the physics
near this fixed point.  In the spin sector, the leading irrelevant 
operator is now the dimension-2 operator 
${\bf J}'_R \cdot {\bf J}'_R(0)$.  Performing an instanton gas 
expansion,\cite{schmid} we find the leading irrelevant operator 
$\cos \sqrt{16\pi} \phi_{R,f}(0)$ in the flavor sector.  Due 
to the Neumann boundary conditions, this operator has dimension 
$2K$ and is irrelevant for $K > 1/2$.  Hence, we see that the 
1-channel Kondo fixed point is stable for $K > 1/2$.

%%%%%%%%%%%%%%%%%%%%%%%%%%%%%%%%%%%%%%%%%%%%%%%%%%%%%%%%%%%%%%%%%%%%%%%%%%%
%%%%%%%%%%%%%%%%%%%%%%%%%%%%%%%%%%%%%%%%%%%%%%%%%%%%%%%%%%%%%%%%%%%%%%%%%%%

Now we consider the more general (and probably more realistic)
case where there is no particle-hole symmetry, and the $V$ and 
$V'$ terms in Eq.~\ref{schreifferwolff} are present.  
Using Eq.~\ref{cosetelectron} and the OPE's of Eqs.~\ref{WZWOPE}
and \ref{IsingOPE}, we obtain 
\begin{eqnarray}
 H_{\rm int} & = & v_{\sigma}~ \lambda_1~ {\bf \tau} \cdot {\bf J}_R(0)
   + \sqrt{v_{\sigma}v_{\rho}}~ \lambda_2~{\bf \tau} \cdot {\bf \phi}~ 
     \cos \sqrt{4\pi} \phi_{R,f}(0)  \nonumber \\
   & + & \sqrt{v_{\sigma}v_{\rho}}~ \lambda_3 ~\epsilon~ 
         \sin \sqrt{4\pi} \phi_{R,f}(0) 
   + v_{\rho}~ \lambda_4~ \partial_x \phi_{R,c}(0)
 \, , \nonumber \\
\label{bosonschreifferwolff2}
\end{eqnarray}
where $\{\lambda_i\}$ are dimensionless couplings with 
$v_{\sigma}~\lambda_1 \sim J$, 
$\sqrt{v_{\sigma}v_{\rho}}~\lambda_2 \sim J'$, 
$\sqrt{v_{\sigma}v_{\rho}}~\lambda_3 \sim V$ 
and $v_{\rho}~\lambda_4 \sim V'$.
To begin with, we can take $\lambda_4$ term into account 
exactly.  Its effect is to produce a small phase shift.  
At the ultraviolet fixed point, $\lambda_1$ and $\lambda_2$ 
satisfy the same RG equations as Eq.~\ref{ultravioletRG};  
$\lambda_3$ satisfies
\begin{equation}
 \frac{d\lambda_3}{dl} = \frac{1}{2}\left(1 
   - \frac{1}{K} \right)\lambda_3  \, , 
\end{equation}
and hence is irrelevant.
Therefore, the flows near the ultraviolet fixed point are
the same as for the particle-hole symmetric case. --- For 
$K \approx 1$, the system flows to the 1-channel Kondo fixed 
point, and for $K$ sufficiently less than unity, the system 
flows toward the 2-channel Kondo fixed point.

Near the 2-channel Kondo fixed point, we still have the 
operators perturbing this fixed point as we did for the 
particle-hole symmetric case.  However, now the $V$ and 
$V'$ terms in Eq.~\ref{schreifferwolff} allow terms which
break particle-hole symmetry.  
Similar to the ultraviolet fixed point, the most relevant 
particle-hole symmetry breaking terms are 
$\partial_x \phi_{R,c}(0)$ of dimension 1 and 
$\epsilon \sin \sqrt{4\pi} \phi_{R,f}(0)$ of dimension 
$(1/2)(1 + 1/K)$.  Therefore, as in the particle-hole 
symmetric case, the 2-channel Kondo fixed point is stable 
for $K < 1/2$; for $K > 1/2$, the 2-channel Kondo fixed point 
is unstable and the system flows to the 1-channel Kondo fixed 
point.  

For $K > 1/2$, the system flows to the 1-channel Kondo 
fixed point.  Now, we must consider the properties of 
the system near this fixed point.  
As we saw for the case of non-interacting electrons, at the 
1-channel Kondo fixed point, the two semi-infinite leads are 
joined into one infinite lead.  For an infinite Luttinger 
liquid, if we allow terms which break particle-hole symmetry, 
it is well known that the most relevant perturbation is 
potential scattering;\cite{kane} in bosonized form, it is 
$\epsilon \sin \sqrt{4\pi} \phi_{R,f}(0)$.  Due to the Neumann 
boundary conditions in the flavor sector, this operator has 
dimension $(1/2)(1 + K)$, and hence is relevant.  This operator 
causes a flow to an insulating fixed point.  Physically, this 
insulating fixed point corresponds to an electron in the 
``even''channel bound in a singlet to the electron on the dot, 
with an infinite potential in the ``odd'' channel.\cite{simon}  
The physics near this fixed point is basically that of two 
semi-infinite Luttinger liquids coupled by weak tunneling.  
The operator for weak tunneling is the most relevant operator 
near this fixed point, and has dimension $(1/2)(1 + 1/K)$;\cite{kane} 
hence, it is irrelevant.  Therefore, this insulating fixed point 
is stable, and is the ultimate fixed point to which the system
flows for $K > 1/2$.

%%%%%%%%%%%%%%%%%%%%%%%%%%%%%%%%%%%%%%%%%%%%%%%%%%%%%%%%%%%%%%%%%%%%%%%%%%%
%%%%%%%%%%%%%%%%%%%%%%%%%%%%%%%%%%%%%%%%%%%%%%%%%%%%%%%%%%%%%%%%%%%%%%%%%%%

\begin{figure}
\centerline{\epsfxsize=1.75in \epsfbox{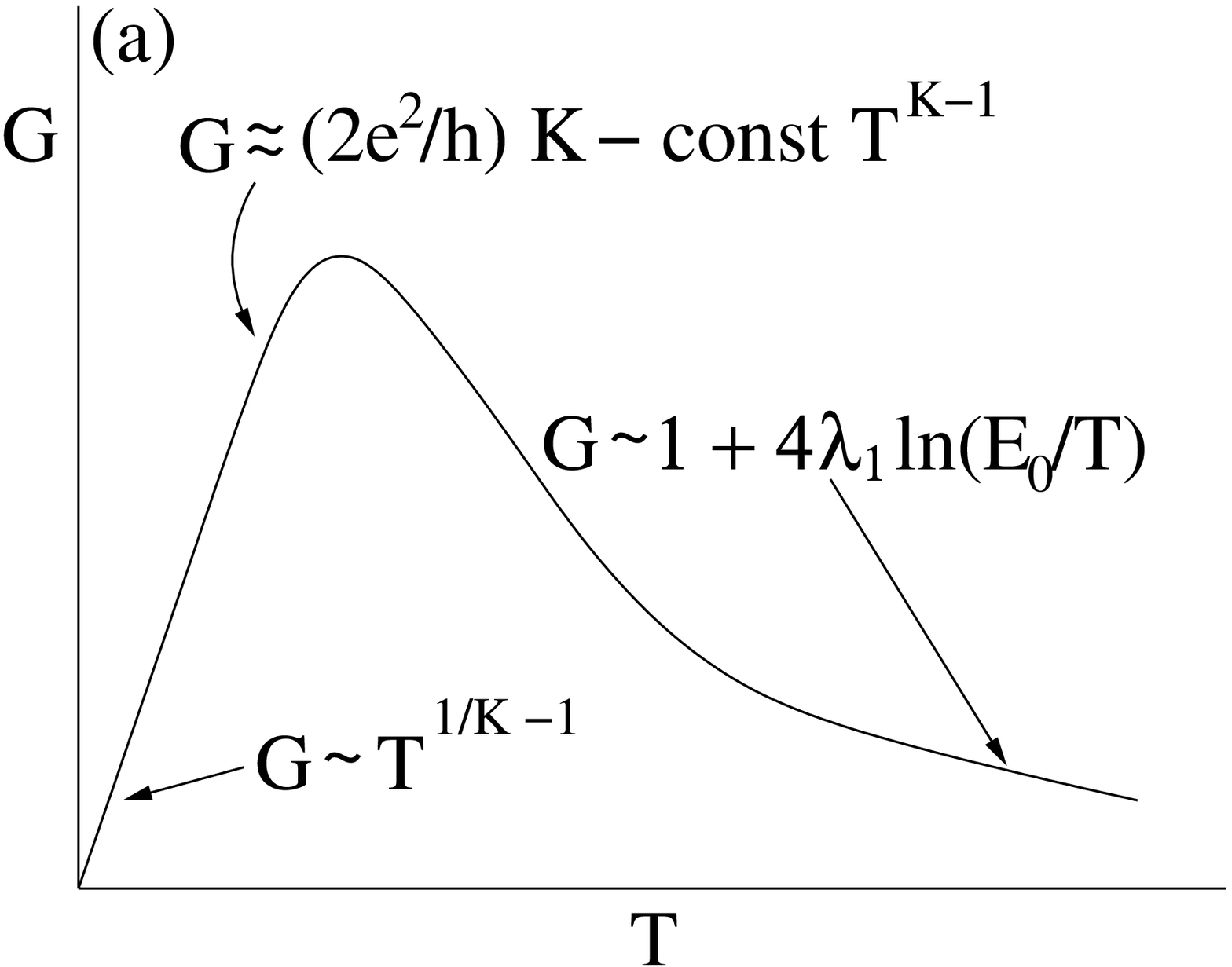} }
\vspace{.2in}
\centerline{ \epsfxsize=1.75in \epsfbox{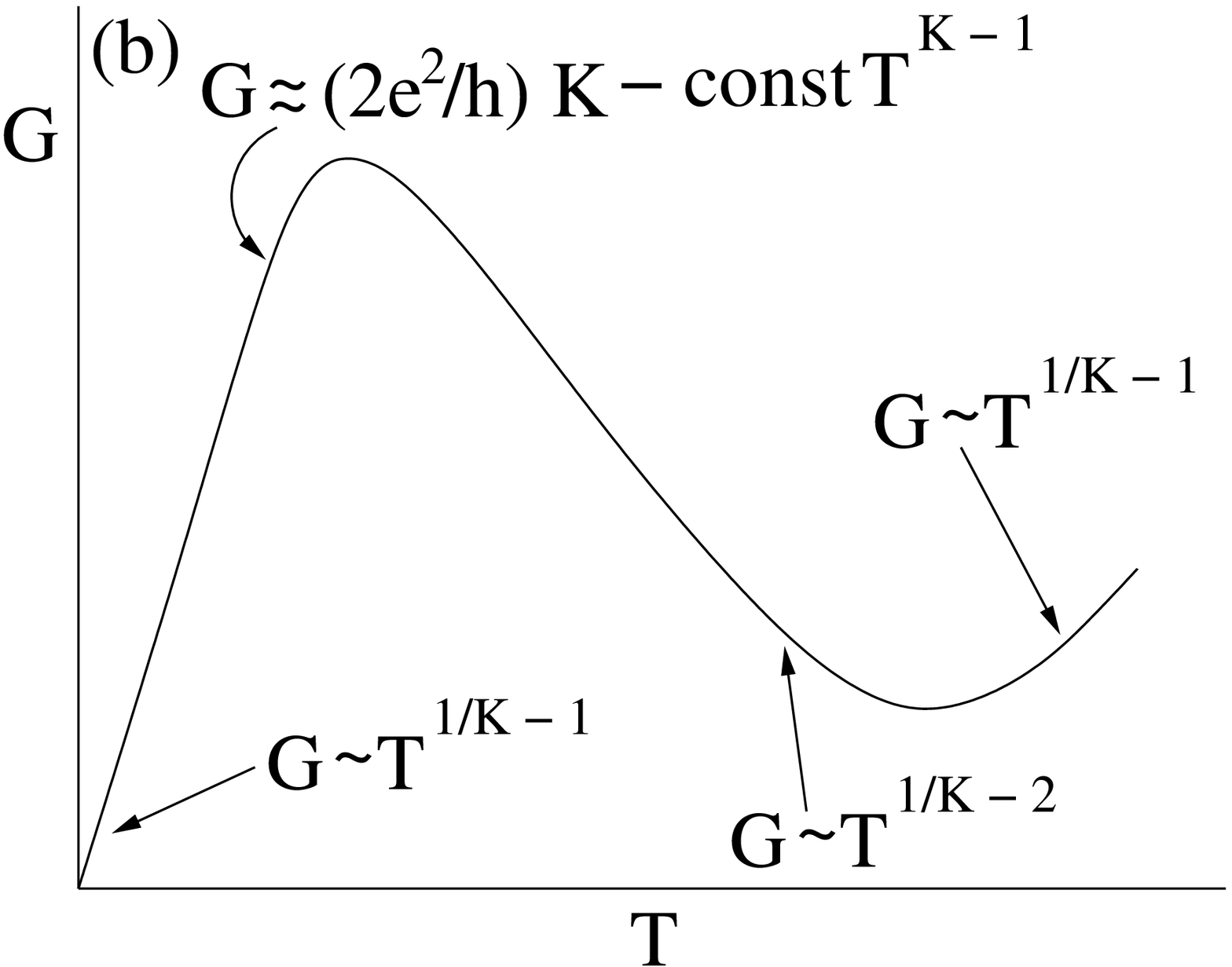}
 \hspace{0.075in} \epsfxsize=1.75in \epsfbox{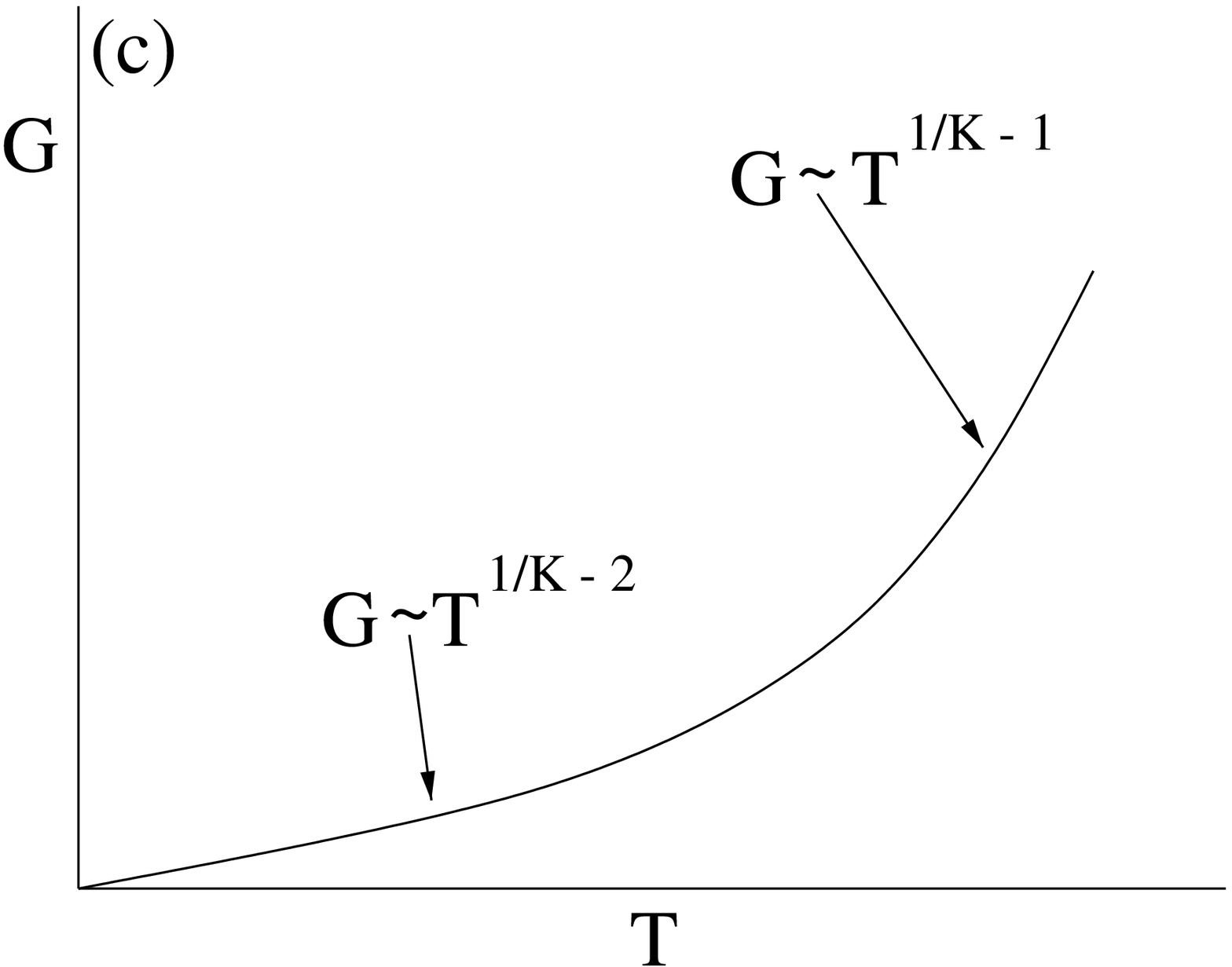} }
\vspace{0.2in}
\caption{ Sketch of the conductance vs. temperature for a 
 quantum dot coupled to Luttinger liquid leads. 
 (a) $K \approx 1$ (b) $K$ sufficiently less than unity, but 
 $K > 1/2$ (c) $K < 1/2$ }
\label{fig:conductance}
\end{figure}

Experimentally, the quantity of interest is the conductance
through the dot, $G$.  From the above discussion, there are 
three distinct cases to consider --- 
{\sl case 1}: $K \approx 1$; {\sl case 2}: $K$ sufficiently less 
than unity, but $K > 1/2$; {\sl case 3}: $K < 1/2$.  Following 
Ref.~\ref{kane}, we compute the conductance perturbatively near
the various fixed points, focussing on the case where the system 
does not have particle-hole symmetry.  The conductance is plotted 
schematically in Fig.~\ref{fig:conductance} for the three cases.

\vspace{.2in}
\vspace{.2in}

\noindent
{\sl case 1}: Near the ultraviolet fixed point, the growth of 
$\lambda_1$ causes $\lambda_2$ to grow logarithmically.  Hence, 
$G$ grows as $\ln(E_0/T)$ ($E_0$ $\sim$ the bandwidth).  As we 
lower the temperature, the system flows to the 1-channel Kondo 
fixed point; $G$ comes close to its maximum value $2e^2 K/h$.  
However, potential scattering is a relevant perturbation to the 
1-channel Kondo fixed point and causes $G$ to decrease as $T^{K-1}$.  
Finally, as $T \rightarrow 0$, $G$ goes to zero as $T^{1/K-1}$. \\
{\sl case 2}: Near the ultraviolet fixed point, $G$ decreases 
as $T^{1/K - 1}$ as the system flows toward the 2-channel Kondo 
fixed point.  However, the 2-channel Kondo fixed point is unstable, 
and $G$ increases as $T^{1/K - 2}$ as we flow away from the 
2-channel Kondo fixed point to the 1-channel Kondo fixed point.  
Near the 1-channel Kondo fixed point, $G$ comes close to its 
maximum value $2e^2 K/h$.  However, potential scattering 
is a relevant perturbation to the 1-channel Kondo fixed point and 
causes $G$ to decrease as $T^{1-K}$.  Finally, as $T \rightarrow 0$, 
$G$ goes to zero as $T^{1-1/K}$. \\
{\sl case 3}: $K_{\rho} < 1/2$ --- Near the ultraviolet fixed
point, $G$ decreases as $T^{1/K - 1}$ as the system flows 
toward the 2-channel Kondo fixed point.  Near the 2-channel 
Kondo fixed point, $G \sim T^{1/K - 2}$.  However, in this 
case the 2-channel Kondo fixed point is stable.  As 
$T \rightarrow 0$, we remain near the 2-channel Kondo fixed
point and $G$ goes to zero as $T^{1/K - 2}$.

In all three cases, $G \rightarrow 0$ as $T \rightarrow 0$.
However, it is important to remember that the fixed points 
to which the system flows are different.  For $K > 1/2$, the 
system ultimately flows to an insulating fixed point of two 
semi-infinite Luttinger liquids coupled by weak tunneling; 
for $K < 1/2$, the system flows to the 2-channel Kondo fixed 
point.  This difference has observable consequences.  In 
particular, if we consider the spin conductance,\cite{kane} 
we find that the 2-channel Kondo fixed point has perfect spin 
conductance, whereas the insulating fixed point has vanishing 
spin conductance.  Though the spin conductance is difficult 
to measure, this effect should be observable in thermal 
conductance measurements.  In particular, we expect a thermal 
conductance, $\kappa$, such that\cite{kane1} 
\begin{equation}
 \frac{\kappa}{T} \rightarrow 
    \left(\frac{\pi^2}{3}\right) \frac{k_B^2}{h} 
    \ \ \ {\rm as} \ \ \ T \rightarrow 0
\end{equation}
if the system flows to the 2-channel Kondo fixed point, whereas 
we expect a thermal conductance such that $\kappa /T$ goes to
zero as $T^{1/K - 1}$ if the system flows to the insulating 
fixed point.

%%%%%%%%%%%%%%%%%%%%%%%%%%%%%%%%%%%%%%%%%%%%%%%%%%%%%%%%%%%%%%%%%%%%%%%%%%%
%%%%%%%%%%%%%%%%%%%%%%%%%%%%%%%%%%%%%%%%%%%%%%%%%%%%%%%%%%%%%%%%%%%%%%%%%%%

To summarize, we considered the Kondo effect in quantum dots
coupled to Luttinger liquid leads, focussing on the case of 
repulsive interactions ($K < 1$) and spin SU(2) symmetry in 
the leads.  We found that the system flows to the 2-channel
Kondo fixed point for $K < 1/2$ and flows to the 1-channel 
Kondo fixed point for $K > 1/2$.  For the particle-hole symmetric 
case, the 1-channel Kondo fixed point is stable; for the 
particle-hole asymmetric case, the system ultimately flows 
to an insulating fixed point.  Furthermore, we computed the 
conductance and found that the qualitative behavior is strongly 
dependent on the value of $K$.  Finally, we pointed out that the 
2-channel Kondo fixed point has perfect spin conductance, which 
should be observable in thermal conductance measurements.

%%%%%%%%%%%%%%%%%%%%%%%%%%%%%%%%%%%%%%%%%%%%%%%%%%%%%%%%%%%%%%%%%%%%%%%%%%%
%%%%%%%%%%%%%%%%%%%%%%%%%%%%%%%%%%%%%%%%%%%%%%%%%%%%%%%%%%%%%%%%%%%%%%%%%%%

This work was supported by the NSERC of Canada.
 
%%%%%%%%%%%%%%%%%%%%%%%%%%%%%%%%%%%%%%%%%%%%%%%%%%%%%%%%%%%%%%%%%%%%%%%%%%%
%%%%%%%%%%%%%%%%%%%%%%%%%%%%%%%%%%%%%%%%%%%%%%%%%%%%%%%%%%%%%%%%%%%%%%%%%%%

%%%%%%%%%%%%%%%%%%%%%%%%%%%%%%%%%%%%%%%%%%%%%%%%%%%%%%%%%%%%%%%%%%%%%%%%%%%
%%%%%%%%%%%%%%%%%%%%%%%%%%%%%%%%%%%%%%%%%%%%%%%%%%%%%%%%%%%%%%%%%%%%%%%%%%%

\end{multicols}

\vspace{-.15in}
  
\begin{thebibliography}{100}
  
\vspace{-.5in}


\bibitem{kondo}{J. Kondo, 
     Prog. Theor. Phys. {\bf 32}, 37 (1964).  }

\bibitem{cox}{For a review, see D.~L. Cox and A. Zawadowski,
    Adv. Phys. {\bf 47}, 599 (1998).  }

\bibitem{dots}{D. Goldhaber-Gordon, {\it et. al.}, 
    Nature {\bf 391}, 156 (1998); 
    S.~M. Cronewett, {\it et. al.}, Science {\bf 281}, 540 (1998);
    F. Simmel, {\it et. al.}, Phys. Rev. Lett. {\bf 83}, 804 (1999);
    W.~G. van der Wiel, {\it et. al.}, Science {\bf 289}, 2105 (2000).}

\bibitem{ng}{T.~K. Ng and P.~A. Lee,
    Phys. Rev. Lett. {\bf 61}, 1768 (1988); 
    L.~I. Glazman and M.~E. Raikh,
    JETP Lett. {\bf 47}, 452 (1988). }

\bibitem{haldane}{F.~D.~M. Haldane, 
    J. Phys. C {\bf 14}, 2585 (1981). }

\bibitem{backscattering}{
Often times, the backscattering term 
$g~{\bf J}_{R,i} \cdot {\bf J}_{L,i}$  
should also be included in Eq.~\ref{leadhamiltonian}.
However, this term is (marginally) irrelevant, and the 
system flows to the fixed point $g=0$.  Though it is 
known to give logarithmic corrections to certain bulk 
correlation functions, it should not affect the leading 
boundary behavior.\cite{frojdh}  Therefore, to simplify 
things, we ignore it.}

\bibitem{eggert}{S. Eggert and I. Affleck,
    Phys. Rev. B {\bf 46}, 10866 (1992).  }

\bibitem{fabrizio}{M. Fabrizio and A.~O. Gogolin,
    Phys. Rev. B {\bf 51}, 17827 (1995). 
    \label{fabrizio}      }

\bibitem{luttingerkondo}{A. Furusaki and N. Nagaosa, 
      Phys. Rev. Lett {\bf 72}, 892 (1994);
 P. Fr\"{o}jdh and H. Johannesson, 
      Phys. Rev. Lett. {\bf 75}, 300 (1995);
 Y.-L. Liu and T.~K. Ng,
      Phys. Rev. B {\bf 61}, 2911 (2000).  
  \label{luttingerkondo}   }
 
\bibitem{goddard}{P. Goddard, A. Kent, and D. Olive,
     Comm. Math. Phys. {\bf 103}, 105 (1986).
     \label{goddard}  }

\bibitem{ludwig2}{I. Affleck, A.~W.~W. Ludwig, and B.~A. Jones,
     Phys. Rev. B {\bf 52}, 9528 (1995).   
     \label{ludwig2}   }

\bibitem{frojdh}{P. Fr\"{o}jdh and H. Johannesson,
     Phys. Rev. B {\bf 53}, 3211 (1996). 
     \label{frojdh}  }

\bibitem{cardy}{J. Cardy, {\it Scaling and
     Renormalization in Statistical Physics}, 
     (Cambridge University Press, Cambridge, 1996). }

\bibitem{ludwig1}{I. Affleck and A.~W.~W. Ludwig,
     Nucl. Phys. B {\bf 360}, 641 (1991).  
     \label{ludwig1}   }

\bibitem{ludwig3}{I. Affleck, A.~W.~W. Ludwig, 
            H.-B. Pang, and D.~L. Cox, 
     Phys. Rev. B {\bf 45}, 7918 (1992).
     \label{ludwig3}    }

\bibitem{blandin}{P. Nozi\`{e}res and A. Blandin,
    J. Phys. (Paris) {\bf 41}, 193 (1980).  }

\bibitem{schmid}{A. Schmid,
     Phys. Rev. Lett. {\bf 51}, 1506 (1983).
     \label{schmid}   }

\bibitem{kane}{C.~L. Kane and M.~P.~A. Fisher, 
     Phys. Rev. B {\bf 46}, 15233 (1992). 
     \label{kane}   }

\bibitem{kane1}{C.~L. Kane and M.~P.~A. Fisher,
     Phys. Rev. Lett. {\bf 76}, 3192 (1996).
     \label{kane1}   }

\bibitem{simon}{P. Simon and I. Affleck,
     cond-mat/0103175.
     \label{simon}   }


\end{thebibliography}
\end{document}